# A Novel Position-based VR Online Shopping Recommendation System based on Optimized Collaborative Filtering Algorithm

Jianze Huang
School of Computer and Data
Engineering
Ningbo Tech University
Zhejiang Province, China

Hao Lan Zhang*
School of Computer and Data
Engineering
Ningbo Tech University
Zhejiang Province, China
haolan.zhang@nit.zju.edu.cn

Huanda Lu
School of Computer and Data
Engineering
Ningbo Tech University
Zhejiang Province, China

Xin Yu
School of Computer and Data
Engineering
Ningbo Tech University
Zhejiang Province, China
line 5: email address or ORCID

Shaoyin Li
School of Computer and Data
Engineering
Ningbo Tech University
Zhejiang Province, China
line 5: email address or ORCID

line 1: 6th Given Name Surname
line 2: *dept. name of organization )*
line 3: *name of organization*
line 4: City, Country
line 5: email address or ORCID

*Abstract*—This paper proposes a VR supermarket with an intelligent recommendation, which consists of three parts. The VR supermarket, the recommendation system, and the database. The VR supermarket provides a 360-degree virtual environment for users to move and interact in the virtual environment through VR devices. The recommendation system will make intelligent recommendations to the target users based on the data in the database. The intelligent recommendation system is developed based on item similarity (ICF), which solves the cold start problem of ICF. This allows VR supermarkets to present real-time recommendations in any situation. It not only makes up for the lack of user perception of item attributes in traditional online shopping systems but also VR Supermarket improves the shopping efficiency of users through the intelligent recommendation system. The application can be extended to enterprise-level applications, which adds new possibilities for users to do VR shopping at home.

## I. INTRODUCTION

Today, virtual reality (VR) has become one of the most promising technology megatrends across digital domains. According to IDC 2021, the augmented global reality and virtual reality (AR/VR) headset market grew 92.1% year over year in 2021. Shipments of AR/VR headsets are forecast to grow 46.9% year-over-year in 2022, reaching double-digit growth by 2026. VR has three main features: 360-degree visual freedom, interactivity, and real-time[1]. 360-degree visual freedom is the key feature of VR[2], which means that users can move their view freely in 360 degrees and experience the immersive feeling. The second is characterized by real-time, which means that objects in the virtual environment are running in real-time to immerse the user. The third feature, interactivity, refers to the extent to which the user can manipulate objects in the simulated environment and the natural degree of feedback received from the environment, i.e., if the user touches an object in the virtual space and moves it, the object's position and state should also change. So far, VR technology has been used in several fields, such as VR conferencing[3], advertising[4], tourism [5] and so on. With the development of technology,

VR devices are not as expensive as they used to be, such as Google's Cardboard, which allows users to interact with objects in the virtual environment through gaze control. In September 2020, Facebook released the Oculus Quest2, which supports wireless streaming mode, allowing users to walk around virtual scenes more freely. More importantly, young people nowadays maintain a strong interest in VR technology. This means that the time is ripe for research and development of immersive VR.

Virtual shopping is the most promising area for VR business model[6-9]; eBay (virtual reality department store), Alibaba (Buy +), IKEA (virtual reality kitchen showroom) and other business giants are trying to embed business services into virtual reality in an attempt to revolutionize the business shopping model again. VR shopping creates an opportunity for users to enter any shopping environment, anytime, anywhere, addressing space and time constraints[10] and improving the overall efficiency of users' shopping. However, this virtual shopping software cannot make personalized recommendations to users. So to overcome this difficulty, a recommendation system is a good choice. Collaborative filtering (CF) is the most widely known technique for this system [11,12]. Collaborative filtering algorithms are divided into User Collaborative Filtering (UCF) and Item Collaborative Filtering (ICF). Both algorithms have a recommendation system model based on the user-item rating matrix. However, the difference is that ICF analyzes the similarity between items and then takes the items with higher similarity to the target user's rating as the result of the recommendation list. At the same time, UCF selects people with similar tastes to the target user in the massive user data and then recommends items to the Recommend items to target users. Related report[13] show that the recommendation system increases sales for Amazon by more than 30%.

The inability of people to gather offline on a large scale due to the impact of the new coronavirus has prompted people to enter the online environment for online shopping. However, the inability of 2D shopping to interact with the products due to its limitations and the lack of user perception of product attributes has led to the inability of online shopping to surpass the user experience of physical stores in this regard. Therefore, this paper designs a VR supermarket software that uses collaborative filtering as a recommendation system. The software enables users to interact with products in a virtual world through VR technology to enhance their shopping experience. It provides personalized recommendations for target users through a recommendation system using collaborative filtering algorithms.

## II. RELATED WORK

### 1. Recommendation System

"Collaborative filtering" was first introduced by Dave Goldberg et al. in 1992[14]. The recommendation of the target user is based on the similarity of the target user's preferences with other users, i.e., (UCF). However, at this time, collaborative filtering still has the following defects.

User-to-user preference similarities (UCF) must be constantly recalculated, and performance is very poor when UCF scale to millions of users。 Therefore, in 2001, Badrul Sarwar et al. pioneered the "item-item" collaborative filtering algorithm, i.e., (ICF)[15]. Since ICF recommends products to users by the similarity between items based on their historical behaviour of selecting items, the similarity calculation requires very little calculation

In order to improve the recommendation performance under cold start conditions, Hyung Jun Ahn [16] proposed a new similarity measure called PIP since the system cannot accurately provide users with satisfactory recommendation results without a large amount of data (cold start). Moreover, by comparing the performance experiments of five measures, COR, COS, CPC, SRC, and PIP, the effectiveness of the PIP similarity measure is demonstrated.

Traditional recommendation systems ignore the intrinsic connection between user preferences and time. Cui et al. [17]proposed a model based on the time correlation coefficient. Using the cuckoo search algorithm, they improved the K-means algorithm (CSK-means) to design an effective personalized recommendation model (PTCCF) based on preference patterns. Extensive experiments were conducted on two real datasets, MovieLens and Douban, and the accuracy of PTCCF was improved by about 5.2% compared with the MCoC model. The effectiveness of the model was demonstrated.

We initially implemented an apriori algorithm recommendation system in python. We packaged it as an exe file, but the algorithm needed to recalculate the strong correlation of the data each time it made a recommendation, which led to performance degradation. Moreover, the recommendation results of the apriori algorithm were poor, with higher confidence and support than some normal values (depending on the strong correlation of the data), and many data did not show any recommendation. In addition, since it is a c# call to exe file, there is some lag when the software runs the recommendation system for the first time (mainly affected by the first c# call to exe file to establish the channel). Therefore, we use a coroutine to establish the channel when the software is started for the first time, implement the collaborative filtering algorithm (ICF) in python language, and propose the ICF-STR algorithm to solve the cold start problem of ICF.

### 2. VR supermarket

VR technology has great potential in business. VR technologies usually use large screen[18] or head-mounted displays to provide users with an immersive sensory experience in real-time. Two classic examples are the CAVE projection system and head-mounted displays. CAVE was first introduced in 1992 by C. Cruz-Neira et al. [19],CAVE is an immersive VR system based on the projection of room sizes. The projector is aligned with three to six projection walls when entering the system. Using a 3D tracker, the user can get close to virtual 3D objects in a system surrounded by projection walls or roam the "real" virtual environment at will. However, the price of the device and the space required for the device are obstacles to the diffusion of such a user experience. Another example is the head-mounted display; the device, through the eye, the output further visual feedback, giving users a sense of 3D experience.

In 2016, eBay partnered with Myer to launch the first virtual reality department store app. eBay's virtual department store app is available on cell phones and can be used with headset devices such as the Samsung Gear VR. Instead of controlling the handle, users gaze at the target product for product selection. Alibaba also released Buy+ in the same year, allowing users to enter the streets of Japan, the United States and Australia and see local stores. Users need to follow prompts to click and buy products in panoramic mode. In VR mode, users put on a VR device and stare at the product for more than 3s to see the product information and make a purchase. In other words, view the product - buy it now - confirm the order - confirm the payment - and pay successfully. However, the downside is that Buy+ cannot recommend products, and eBay's simple category recommendations cannot meet the needs of increasingly sophisticated users.

Therefore, this paper embeds the improved collaborative filtering recommendation algorithm into the VR supermarket as the recommendation system of this supermarket. This VR supermarket will make personalized recommendations according to different users' demands for products, thus improving users' shopping experience.

## III. PROPOSED DESIGN

The VR program system is shown in Figure 1. The system consists of the VR application, the recommendation system, and the external database.

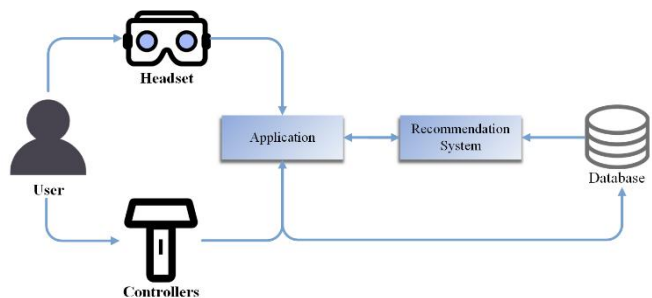

Figure 1 Software design flow chart

As shown in Figure 1, when a user starts to enter the VR supermarket system and picks up an item, the user will see the information panel for that item. If the user does not purchase the item, he/she can choose to check out directly, and the information about the purchased item will be added to the database. If the user purchases the item, the recommendation system will use the item information in the database to calculate the similarity between the item just purchased and other items, as well as the similarity between the items already purchased in the shopping cart (combined) and other items, and combine the user's evaluation to predict the items the user may purchase, and finally recommend them to the user in the form of a recommendation panel.

The Unity game engine and input develop the VR supermarket system through VR devices. The system is designed with an interface for passing information data to the recommendation system. The recommender system analyzes the data provided by the VR supermarket system based on the data from the external database. It transmits the analyzed data back to the VR supermarket system, and the user can get the data feedback from the recommender system through the VR headset. As the recommendation system (ICF) needs to analyze from two dimensions, one is the similarity between different goods and the other is the user's evaluation. The recommendation system can use the cosine similarity formula to calculate the similarity between products. For the user's evaluation system, the following scoring rules are designed in this paper. As shown in Table 1.

Table 1 Scoring Judgment Criteria

| After viewing the panel results | Extra credit |
| --- | --- |
| Purchase of goods | +1 |
| No purchase of goods | -1 |
| Buy Recommended Products | +1.5 |
| Do not buy recommended products | -0.5 |

## IV. Method

### 1. VR supermarket and recommendation system

In this paper, we use the Unity game development engine and SteamVR (plug-in) to develop a VR supermarket together and use HTC Vive as the device to experience the VR supermarket. The experience flow of the user wearing the device is shown in Figure 2.

### 2. Recommended System Improvements

A recommender system is often defined as a system that can provide intelligent recommendations based on the user's preferences and interests. At VR Supermarket, we have three approaches that can be used as algorithms for recommendation systems. They are the apriori algorithm, the user-based UCF algorithm, and the product-based ICF algorithm. In order to verify which of these three algorithms is more suitable for the recommendation system, we used the dataset of MovieLens as the dataset of the recommendation system and conducted an experimental comparison. The Apriori algorithm is formulated as follows:

$$D(x,y) = \left( \sum_{u=1}^{n} (x_u - y_u)^p \right)^{\frac{1}{p}} \qquad (1)$$

The x and y in the formula represent the two products for which product similarity is to be calculated, respectively. When p=2, the Euclidean distance (Euclidean distance) distance is obtained, which is the straight-line distance between two points (hereafter referred to as Euclidean distance). Each feature parameter in the Euclidean distance is equally weighted.

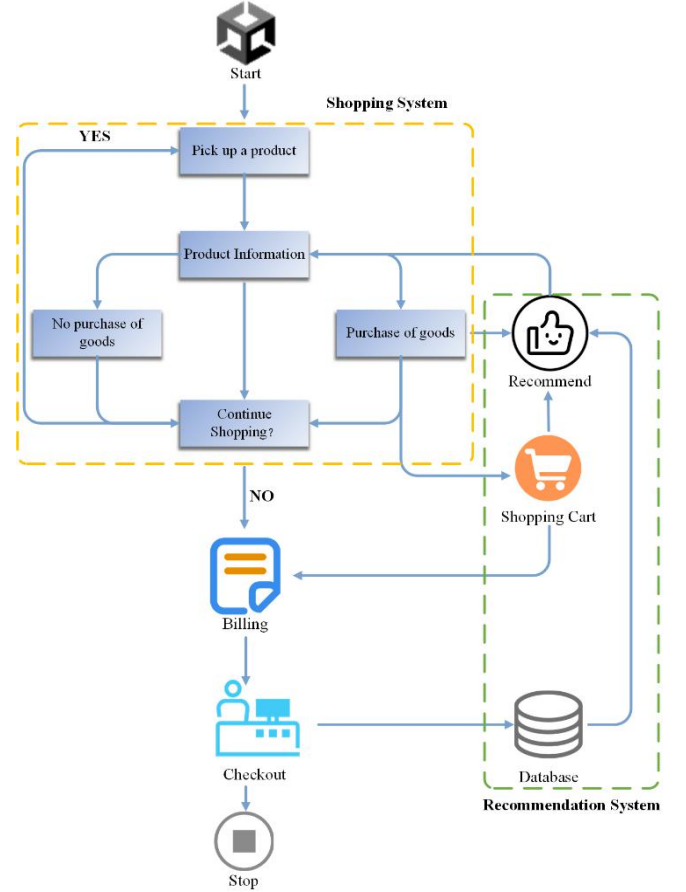

Figure 2 User Flow Chart.

In order to choose the most suitable recommendation algorithm, we not only experimented with the Apriori algorithm but also with ICF and UCF. The core of the ICF algorithm and UCF algorithm is to calculate user similarity and product similarity, respectively. The common methods to calculate similarity include Pearson correlation coefficient, cosine similarity, and Euclidean distance. Since the accuracy and practicality of cosine similarity recommendation are relatively high, this paper adopts the cosine similarity formula to calculate the similarity between products. The formula for calculating the cosine similarity is as follows:

$$sim(x,y) = cos\theta = \frac{\overline{xy}}{||x|| \cdot ||y||} = \frac{\sum_{i=1}^{n} x_i y_i}{\sqrt{\sum_{i=1}^{n}(x_i)^2} \sqrt{\sum_{i=1}^{n}(y_i)^2}} \qquad (2)$$

The x and y in the formula represent the two products or users for which similarity is to be calculated. The larger the result calculated by the cosine similarity formula, the more similar these two products or users are.

The apriori algorithm requires strongly correlated data to make recommendations, so in practical situations, this algorithm does not recommend well. Although the UCF algorithm improves the disadvantage of the apriori algorithm that requires strongly correlated data, this UCF algorithm is only suitable for computing a small number of users, and the cost of maintaining the user similarity matrix increases as the number of users increases, so we finally adopted the ICF algorithm.

The formula (2) shows that the traditional cosine similarity requires data from user feedback to calculate the product similarity. When the product data is insufficient or the user rating is low, it will lead to the "cold start" problem of the recommendation system. In this paper, we propose two solutions to this problem, which are used to ensure that the recommendation interface appears every time a user buys a product.

1) When a user buys an item directly without stopping at any shelf and does not give any recommendation list, the recommendation system generates several random numbers to recommend random items in the supermarket.

2) When the user stays on one or more shelves for more than 10s and does not buy any goods on this shelf, the system will record the information of the goods that the user did not buy on this shelf. When more than one shelf information meets this requirement, the system will only take the most recent shelf information as a recommendation to buy the next cold start problem goods. The shelf specific stay range refers to the following chart:

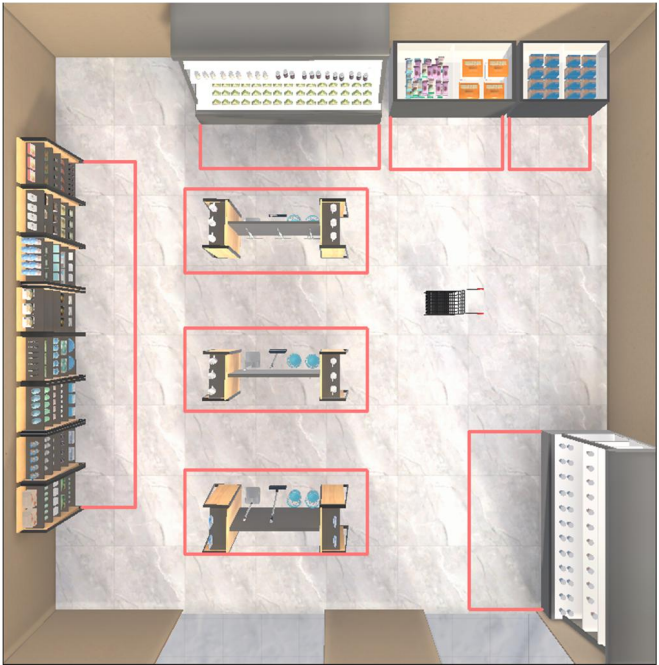

Figure 3 Area determination map

The red range in the image indicates that the system determines that the user is standing in front of this shelf.

## V. RESULTS AND DISCUSSION

To verify the effectiveness of the improved ICF algorithm as a recommendation system algorithm, we used the dataset from MovieLens. Since the dataset on MovieLens is a dataset with a rating nature, it can be generalized to the dataset collected by VR supermarkets as a way to demonstrate that the improved ICF algorithm can be used as a recommendation algorithm for VR supermarket recommendation systems. The dataset on MovieLens is quite large, with the training set consisting of 800168 user-rated movie records and the test set consisting of 200042 user-rated movie records. We will verify the effectiveness of the improved ICF algorithm as an algorithm for the VR supermarket recommendation system from the following three experiments. The CPU used in this paper is i7-11800H.
1. Recommended rate of the algorithm:

| Algorithm | Recommended speed |
|---|---|
| Apriori(min_support=0.7) | 0.21s |
| Apriori(min_support=0.15) | 7.45s |
| ICF | 0.0029s |
| UCF | 0.0019s |
| ICF-STR | 0.0031s |

Table 2 Speed Comparison

Min_support represents the minimum support of the Apriori algorithm. According to Table II, the Apriori algorithm runs the slowest at a support degree of 0.15. Although the Apriori algorithm runs fast at a support degree of 0.7, there is no strongly correlated data, i.e., no product recommendations appear. Therefore, the Apriori algorithm is not suitable for the recommendation system of VR supermarket.
2. Model training speed：

| Algorithm | Training rate |
|---|---|
| ICF | 180.25s |
| UCF | 381.27s |
| ICF-STR | 180.25s |

Table 3 Model Rate Comparison

The dataset used in this paper has 3952 movies and 6040 user ratings. As can be seen from Table 3, the model training rate of the UCF algorithm is significantly lower than that of the ICF algorithm. This is because the time complexity of UCF is approximately equal to the square of the number of users, and when the number of users becomes larger, it becomes increasingly difficult to calculate the similarity between users. Since the VR supermarket has to recommend in real-time if the user is a new user, the system has to recalculate the similarity between each user and the new user for each new item purchased, and the amount of calculation will keep increasing as the number of users increases, so the UCF algorithm is not suitable for the recommendation system of VR supermarket.
3. Recommendations under Cold Start

We packaged the algorithms as exe and called them in VR supermarket. When the system has no data set or only a small amount of data set, the results of each recommendation algorithm are shown in the table below：
1. Did not stop at any shelf

| Algorithm | Recommended Results |
|---|---|
| Apriori | NULL |
| ICF | NULL |
| UCF | NULL |
| ICF-STR | Showergel |

Table 4    Random recommendation results

According to Table 4, traditional recommendation algorithms cannot recommend any product without a dataset.

In this case, ICF-STR solves this drawback. When a user picks up a product directly without stopping at any shelf, the system will make a random recommendation.

2. Staying in front of the shelf:

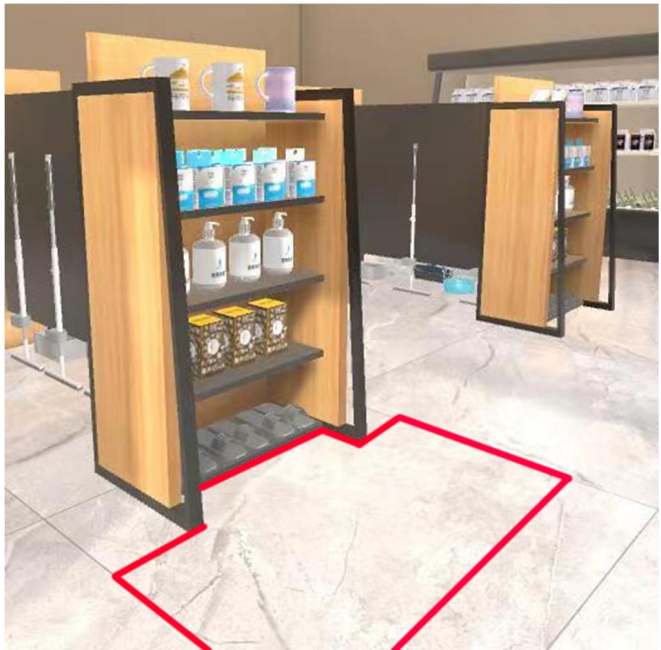

Figure 4 The user stands on the shelf

As in Figure 4, when the user stands in the red area and stays for more than 10 s. The system will use the products on this shelf as the recommended content. The specific results are as follows：

| Algorithm | Recommended Results |
|---|---|
| Apriori | NULL |
| ICF | NULL |
| UCF | NULL |
| ICF-STR | Mug, band-aid, disinfectant, coffee, slippers |

Table 5 Shelf recommendation

As can be seen from Table 5, only ICF-ST appears with specific recommendations, and the recommendations are what is on the shelf. This approach can recommend products to users more effectively than random recommendations.

The result of running in VR Supermarket is as follows:

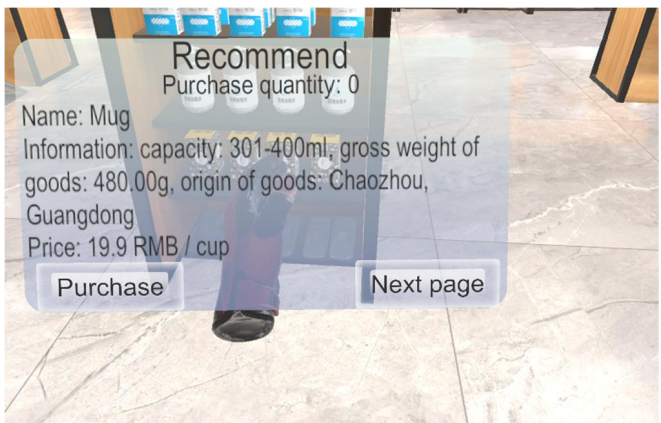

Figure 5 VR supermarket recommended results

As shown in Figure 5, when the recommendation panel appears, users can see the basic information of the recommended items, as well as the purchase situation, click to buy, the recommended items will be added to the shopping cart, and click the next page users can view other recommended items.

## VI. CONCLUSION

The traditional online shopping model, where users are unable to interact with the products, as well as the lack of perception of product attributes, results in online shopping being unable to surpass the user experience of physical stores in these aspects. VR supermarkets can make up for this deficiency and increase the shopping experience of users. An intelligent recommendation system, which can effectively recommend specific products to target users, improves their purchasing efficiency. Thus the VR supermarket system has great potential compared to traditional purchasing models. VR supermarkets can be customized for different sizes of supermarkets and or specific needs. On this basis, it can be combined with voice narration or NPC interaction to further enhance the user's shopping experience.

## ACKNOWLEDGMENT

This work is partially supported by Humanity and Social Science Foundation of the Ministry of Education of China (21A13022003), Zhejiang Provincial Natural Science Fund (LY19F030010), Zhejiang Provincial Social Science Fund (20NDJC216YB), Zhejiang Provincial Educational Science Scheme 2021 (GH2021642) and National Natural Science Foundation of China Grant (No. 72071049), Ningbo public welfare science and technology plan Grant No. 2021S093.

## REFERENCES

[1] Y. Xing, Z. Liang, J. Shell, C. Fahy, K. Guan, and B. Liu, "Historical Data Trend Analysis in Extended Reality Education Field," in *2021 IEEE 7th International Conference on Virtual Reality (ICVR)*, 2021, pp. 434-440.

[2] M. Speicher, B. D. Hall, and M. Nebeling, "What is mixed reality?," in *Proceedings of the 2019 CHI conference on human factors in computing systems*, 2019, pp. 1-15.

[3] S. N. B. Gunkel, M. D. W. Dohmen, H. Stokking, and O. Niamut, "360-Degree Photo-realistic VR Conferencing," *2019 IEEE Conference on Virtual Reality and 3D User Interfaces (VR)*, pp. 946-947, 23-27 March 2019.

[4] L. De Gauquier, M. Brengman, K. Willems, and H. Van Kerrebroeck, "Leveraging advertising to a higher dimension: experimental research on the impact of virtual reality on brand personality impressions," *Virtual Reality*, vol. 23, no. 3, pp. 235-253, Sep 2019.

[5] D. A. Guttentag, "Virtual reality: Applications and implications for tourism," *Tourism Management*, vol. 31, no. 5, pp. 637-651, 2010/10/01/ 2010.

[6] K. Cowan and S. Ketron, "A dual model of product involvement for effective virtual reality: The roles of imagination, co-creation, telepresence, and interactivity," *Journal of Business Research*, vol. 100, pp. 483-492, 2019/07/01/ 2019.

[7] V. K. Ketoma, P. Schäfer, and G. Meixner, "Development and evaluation of a virtual reality grocery shopping application using a multi-Kinect walking-in-place approach," in *International Conference on Intelligent Human Systems Integration*, 2018, pp. 368-374: Springer.

[8] A. Moes and H. v. Vliet, "The online appeal of the physical shop: How a physical store can benefit from a virtual representation," *Heliyon*, vol. 3, no. 6, p. e00336, 2017/06/01/ 2017.

[9] E. Sikström, E. R. Høeg, L. Mangano, N. C. Nilsson, A. De Götzen, and S. Serafin, "Shop'til you hear it drop: influence of interactive auditory feedback in a virtual reality supermarket," in *Proceedings of the 22nd ACM Conference on Virtual Reality Software and Technology*, 2016, pp. 355-356.

[10] B. Serrano, R. M. Baños, and C. Botella, "Virtual reality and stimulation of touch and smell for inducing relaxation: A randomized controlled trial," *Computers in Human Behavior*, vol. 55, pp. 1-8, 2016/02/01/ 2016.

[11] J. Bu, X. Shen, B. Xu, C. Chen, X. He, and D. Cai, "Improving Collaborative Recommendation via User-Item Subgroups," *IEEE Transactions on Knowledge and Data Engineering*, vol. 28, no. 9, pp. 2363-2375, 2016.

[12] W. S. Hwang, J. Parc, S. W. Kim, J. Lee, and D. Lee, ""Told you i didn't like it": Exploiting uninteresting items for effective collaborative filtering," in *2016 IEEE 32nd International Conference on Data Engineering (ICDE)*, 2016, pp. 349-360.

[13] B. Smith and G. Linden, "Two Decades of Recommender Systems at Amazon.com," *IEEE Internet Computing*, vol. 21, no. 3, pp. 12-18, 2017.

[14] D. Goldberg, D. Nichols, B. M. Oki, and D. Terry, "Using collaborative filtering to weave an information tapestry," *Communications of the Acm*, vol. 35, no. 12, pp. 61-70, Dec 1992.

[15] B. Sarwar, G. Karypis, J. Konstan, and J. Riedl, "Item-based collaborative filtering recommendation algorithms," in *Proceedings of the 10th international conference on World Wide Web*, 2001, pp. 285-295.

[16] H. J. Ahn, "A new similarity measure for collaborative filtering to alleviate the new user cold-starting problem," *Information Sciences*, vol. 178, no. 1, pp. 37-51, 2008/01/02/ 2008.

[17] Z. Cui *et al.*, "Personalized Recommendation System Based on Collaborative Filtering for IoT Scenarios," *Ieee Transactions on Services Computing*, vol. 13, no. 4, pp. 685-695, Jul-Aug 2020.

[18] N. Zikic, "Evaluating relative impact of VR components screen size, stereoscopy and field of view on spatial comprehension and presence in architecture," Citeseer, 2007.

[19] C. Cruz-Neira, D. J. Sandin, T. A. DeFanti, R. V. Kenyon, and J. C. J. C. o. t. A. Hart, "The CAVE: audio visual experience automatic virtual environment," vol. 35, no. 6, pp. 64-73, 1992.